\documentclass[letterpaper,11pt]{article}
\usepackage{ArtStyShr} 
\bibliography{LnSbSpOCT}
\copyrightyear{2020}

\title{\vspace{-5ex} \LARGE \bf Shot-noise limited, supercontinuum based optical coherence tomography}
\author[1,*]{Shreesha~Rao~D.~S.}
\author[1]{Mikkel~Jensen}
\author[1]{Lars~Gr\"{u}ner-Nielsen}
\author[2]{Jesper~Toft~Olsen}
\author[3]{Peter~Heiduschka}
\author[4]{Bj\"{o}rn~Kemper}
\author[4]{J\"{u}rgen~Schnekenburger}
\author[5]{Martin~Glud}
\author[5]{Mette~Mogensen}
\author[1]{Niels~M\o ller~Israelsen}
\author[1,2]{Ole~Bang}

\affil[1]{\footnotesize DTU Fotonik, Dept. of Photonics Engineering, Technical University of Denmark, \O rsteds Plads, Kongens Lyngby, 2800, Denmark.}
\affil[2]{\footnotesize NKT Photonics A/S, Blokken 84, 3460 Birker\o d, Denmark.}
\affil[3]{\footnotesize Department of Ophthalmology, University of M\"{u}nster Medical Centre, Domagkstr. 15, M\"{u}nster, D-48149, Germany.}
\affil[4]{\footnotesize Biomedical Technology Center of the Medical Faculty, University of M\"{u}nster, Mendelstr. 17, M\"{u}nster, D-48149, Germany.}
\affil[5]{\footnotesize Department of Dermatology, Bisbebjerg Hospital, University of Copenhagen, Bispebjerg Bakke 23, Copenhagen NV, 2400, Denmark.}

\affil[*]{shdes@fotonik.dtu.dk; shreesharaods@gmail.com.}
\date{}                     

\begin{document}
\maketitle
\begin{abstract}
We present the first demonstration of shot-noise limited supercontinuum-based spectral domain optical coherence tomography (SD-OCT) with axial resolution of 5.9 $\mu$m at a center wavelength of 1370 nm. Current supercontinuum-based SD-OCT systems cannot be operated in the shot-noise limited detection regime because of severe pulse-to-pulse relative intensity noise of the supercontinuum source. To overcome this disadvantage we have developed a low-noise supercontinuum source based on an all-normal dispersion (ANDi) fiber, pumped by a femtosecond laser. The noise performance of our 90 MHz ANDi supercontinuum source is compared to that of two commercial sources operating at 80 and 320 MHz repetition rate. We show that the low noise of the ANDi supercontinuum source improves the OCT images significantly in terms of both higher contrast, better sensitivity, and improved penetration. From SD-OCT imaging of skin, retina, and multi-layer stacks we conclude that supercontinuum-based SD-OCT can enter the domain of shot-noise limited detection.
\end{abstract}
\section*{INTRODUCTION}
Optical coherence tomography (OCT) relies on white light interferometry to non-invasively image translucent samples~\cite{Huang91Sci}. Since its invention nearly three decades ago, OCT has undergone an exceptional technological development which has seen imaging speeds increase more than a factor of a million~\cite{Wie10Mul}, sensitivities improve by 15 dB~\cite{Dre99InVi}, and the resolution enhanced to single cell levels~\cite{Dub18LiFi}. These improvements in speed and sensitivity are largely down to the shift from the original time-domain (TD-) OCT to Fourier domain (FD-) OCT~\cite{deBo17T5}. In TD-OCT, a single photodiode detects the interferometric signal while a mirror is scanning the reference arm of the interferometer. In FD-OCT however, the spectral components of the white light source are detected independently, either with a spectrometer [termed spectral domain (SD-) OCT] or with a wavelength-swept laser and a fast photodiode [termed swept source (SS-) OCT]. A Fourier transform builds an axial scan (A-scan) similar to that created by scanning in TD-OCT. The advantage of both Fourier domain methods over TD-OCT is two-fold: Firstly, eliminating the need for a scanning mirror greatly increases detection speed. Secondly, the independent detection of the spectral components improves the sensitivity~\cite{Leit03FdTd}. This technological advancement has facilitated the widespread clinical use of high-resolution OCT in ophthalmology~\cite{Ho11Doc} as well as the extensive and very active research in other medical fields such as dermatology~\cite{Niel18Val}, oncology~\cite{Yao17Vis} and gastro-enterology~\cite{Ash19Ast}.

Since OCT relies on white light interferometry, the axial resolution ($\Delta$z) of an OCT system is equal to the coherence length of the employed light source. The coherence length, $l_c$, scales proportionally with the square of the centre wavelength of the light source, $\lambda_0$, and inversely proportional with the bandwidth, $\Delta \lambda$. For a light source with a Gaussian spectrum, we have $l_c = \Delta z = (2\ln{2}/\pi)\times(\lambda_0^2/\Delta \lambda$). Reducing the center wavelength to improve the resolution affects both the absorption and scattering of the light in the sample and may introduce unwanted effects such as heating or reduced penetration. A more controllable approach to improve the resolution is therefore to chose a light source with a wide bandwidth.

\begin{figure}[htbp!]
\centering
\fbox{\includegraphics[width=0.9\linewidth]{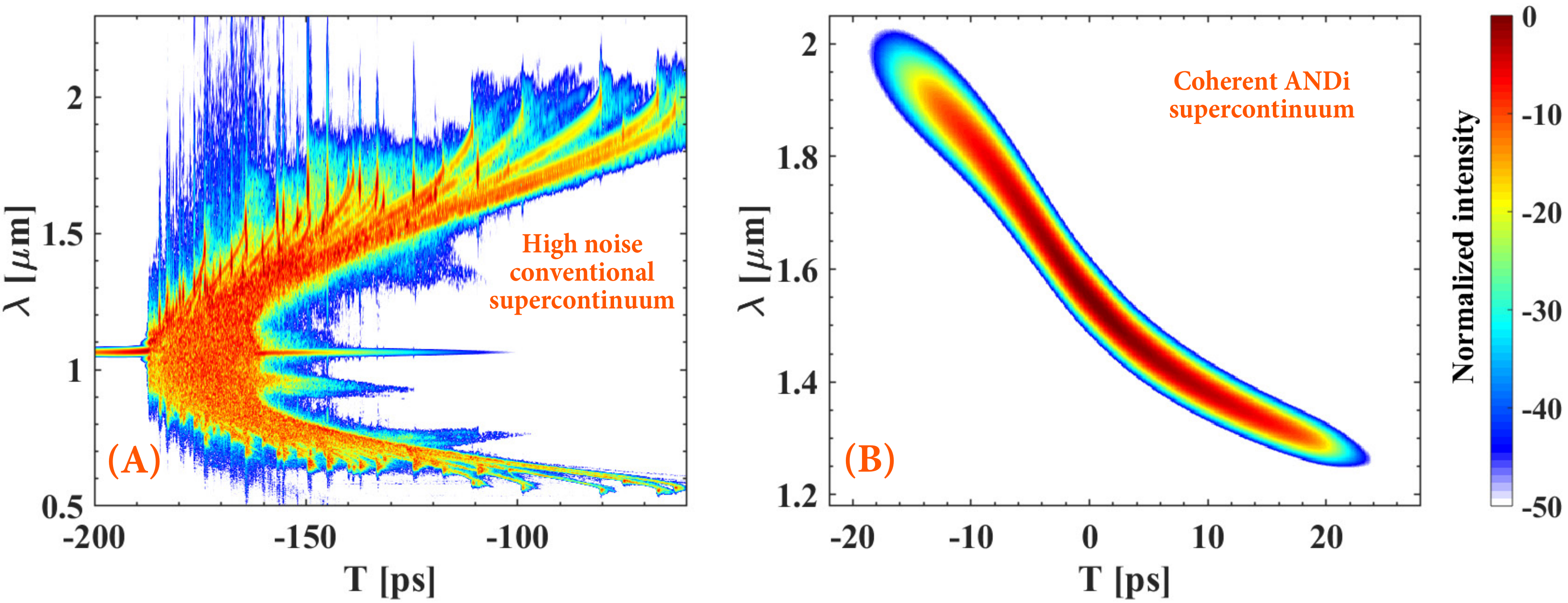}}
\caption{Single-shot output spectrograms of {\bfseries(A)} a typical conventional long-pulse pumped SC source and {\bfseries(B)} a typical fully coherent ANDi SC source}
\label{Fig:Spgm}
\end{figure}

Emerging commercially over the last 15 years, supercontinuum (SC) sources have had an instant impact in the life sciences. An SC based confocal microscope was voted as a top 10 innovation in 2008~\cite{Inn08Sci}. SC sources have also been an increasingly popular choice in OCT when ultra-high axial resolution is the key metric. This is due to their multi-octave spanning spectra, small footprints, robustness, and long-term stability. However, the main disadvantage of SC sources in OCT is the severe intensity noise, which limits the sensitivity of OCT systems due to a noise level far from the ideal shot-noise limit~\cite{Jen19Noi}. Current state-of the art SD-OCT systems use high (320 MHz) repetition rate SC sources to average out some of the intensity noise. The intensity noise is common for all conventional SC sources, and the reason is found in the underlying physics: In most commercial SC sources, the broadband light is generated by launching a long [picosecond (ps) or nanosecond (ns)] light pulse through a nonlinear optical fiber. Complex interplay between linear and nonlinear effects create solitons and dispersive waves (DWs) that constitute an ultra-wide spectrum covering several octaves, as seen in the typical output spectrogram shown in Fig.~\ref{Fig:Spgm}(A). In the regime of long pulses, the initial broadening of the SC is generated by nonlinear amplification of quantum noise~\cite{Dud06Rev}, and the resulting spectra are thus particularly noisy and uncorrelated from pulse-to-pulse. The large integration time needed to reduce the effect of intensity noise to an acceptable level when using conventional SC sources, makes the OCT system unsuitable to image moving objects, such as a flickering eye in ophthalmology. Hence, there is a pressing need for SC sources without any excess noise. The short pulse [femtosecond (fs)] regime has long promised ultra-low noise SC from coherent spectral broadening~\cite{Heidt10Pulp,Heidt11OcC,Klim16Dft2} [typical output spectrogram shown in Fig.~\ref{Fig:Spgm}(B)], but polarization \cite{Ivn18PMI} and nonlinear effects~\cite{Heidt17Lim}, and  pump laser noise \cite{GenAmNo19} introduce severe restrictions on the combination of pulse peak power, pulse length, and fiber length that guarantees low-noise behaviour, making short-pulsed SC sources less broadband and with lower power than their long-pulsed counterparts. As a result, fs-based SC sources are only just emerging in the commercial market. Even research systems have hardly been used in OCT, with the exception of Nishizawa et al.~\cite{Kaw16Fr} who demonstrated SD-OCT with an SC source utilising short pump pulses. While demonstrating imaging of a mouse brain with good sensitivity, Nishizawa et al. did not quantify the noise performance of the SC source or analyze the performance in OCT imaging, and thus it has yet to be demonstrated that shot-noise limited OCT can be achieved with an SC source.  

To answer this fundamental question and for the first time demonstrate that shot-noise limited SC-based SD-OCT is indeed possible, we have developed an SD-OCT system operating around 1370 nm, which uses a low-noise SC source specifically designed for this application. The low-noise SC source uses a fiber with weak all-normal dispersion (ANDi) pumped by a fs laser, in which case the spectral broadening is dominated by the coherent processes of self-phase modulation (SPM) and optical wave-breaking (OWB)~\cite{TakaFirLnS06,Finot08Ben}. We ensured that the normal dispersion is weak and the pulse length is short, in order to specifically avoid noise-generating effects, such as parametric Raman noise~\cite{Heidt17Lim} and polarization mode instability (PMI)~\cite{Ivn18PMI}.

We first compare, one-to-one, the noise characteristics of the OCT system using both commercially available SC sources, SuperK extremes operating at 80 MHz and 320 MHz, as well as our ANDi fiber based low-noise SC source operating at 90 MHz, from now on referred to as just the ANDi SC source. The 320 MHz SuperK extreme source represents today's state-of-the-art in SC sources for SD-OCT due to its high repetition rate. We use all three sources to image a tape-layer phantom, healthy skin, an ex-vivo mouse retina, and ex-vivo fat tissue and compare the images. We demonstrate that in all cases the  SD-OCT system with the ANDi fiber based low-noise SC source provides superior and shot-noise limited performance. 

\section*{RESULTS}
\subsection*{Noise characteristics} 
\begin{figure}[t!]
\centering
\fbox{\includegraphics[width=0.95\linewidth]{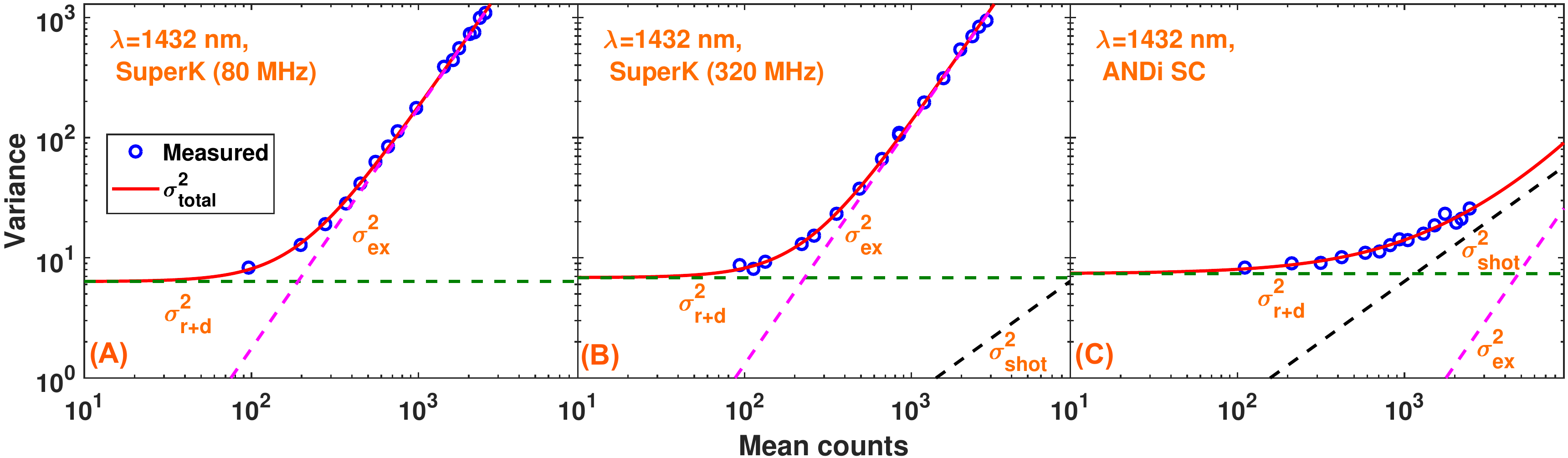}}
\fbox{\includegraphics[width=0.95\linewidth]{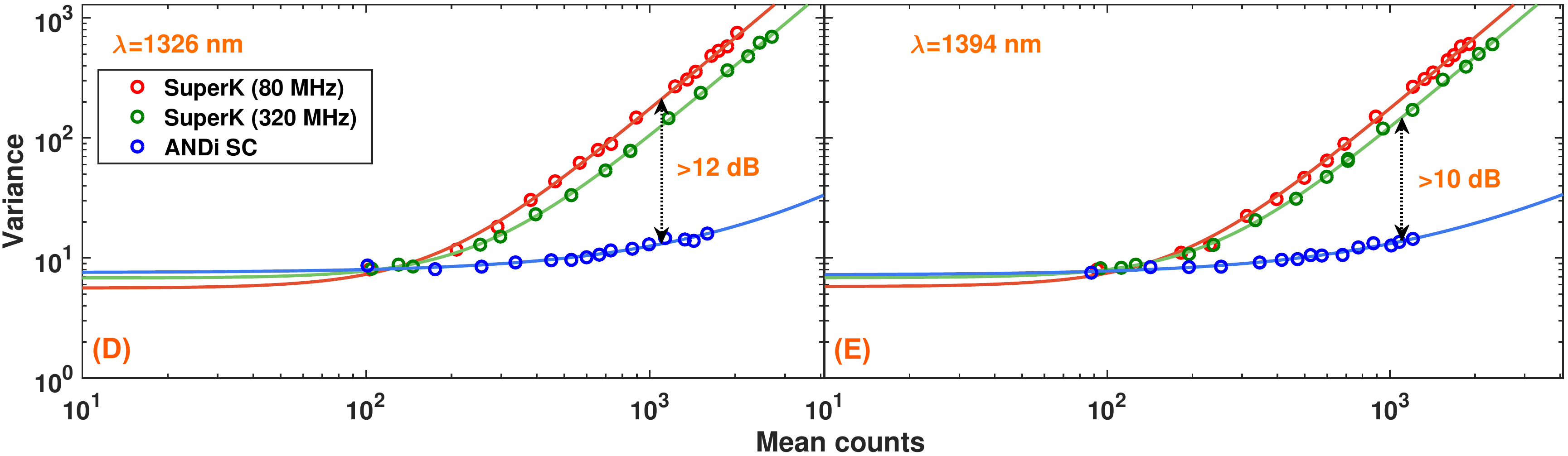}}
\fbox{\includegraphics[width=0.95\linewidth]{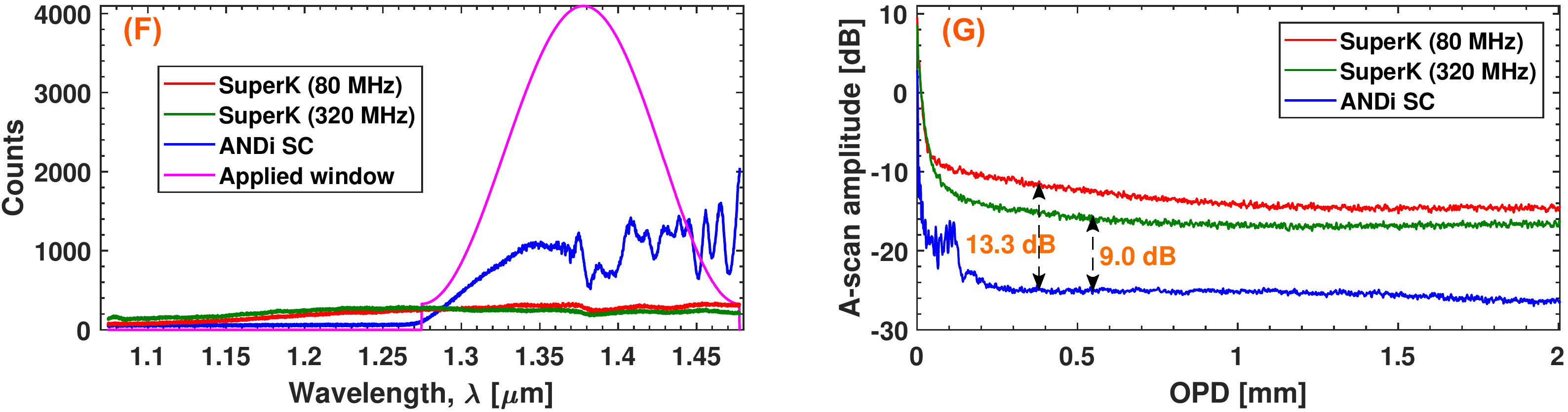}}
\caption{The variance in counts versus the mean counts (over 1024 measurements) at 1432 nm for the 80 MHz SuperK extreme {\bfseries(A)}, 320 MHz SuperK extreme {\bfseries(B)}, and ANDi SC source with 90 MHz repetition rate {\bfseries(C)}. Comparison of the variance in counts versus the mean counts for the 80 MHz SuperK extreme, 320 MHz SuperK extreme, and ANDi SC source with 90 MHz repetition rate at 1326 nm  {\bfseries(D)}, and 1394 nm {\bfseries(E)}. Spectral counts across the spectrometer bandwidth for the three sources, when the counts at 1432 nm is fixed at the optimum value for imaging (250 counts for SuperK extreme sources and 1000 counts for the ANDi SC source) {\bfseries(F)}. The magenta curve shows the window used in the signal processing. {\bfseries(G)} Corresponding variance in A-scans over 1024 measurements. In {\bfseries(D-G)} the 80 MHz SuperK extreme, 320 MHz SuperK extreme, and ANDi SC source are shown as red, green, and blue curves, respectively. All measurements are taken with blocked sample arm.}
\label{Fig:NsMsLnSk}
\end{figure}

The SD-OCT setup is made of a Michelson interferometer [see Fig.~\ref{Fig:SpecARin}(A) in the materials and methods section for further details]. Noise characterization of the OCT system was done with each of the three sources in the wavelength range from 1280-1478 nm, with identical spectrometer operation. The sample arm was blocked and the photo-counts measured in the spectrometer were recorded for various reference arm power levels. The noise of a supercontinuum is known to be wavelength dependent, so the noise of the OCT system will also depend on wavelength or pixel number.
 Figures~\ref{Fig:NsMsLnSk}(A-C) show a representative example at 1432 nm, for the variance of the three sources as a function of mean counts (averaged over 1024 measurements). The circles are the measured values and the continuous line is the fit for the total variance, $\sigma_{tot}^2=\sigma_{r+d}^2+\sigma_{shot}^2+\sigma_{ex}^2$. Here $\sigma_{r+d}^2$ is the read out and dark noise from the spectrometer, which is independent of input power. $\sigma_{shot}^2$ is the shot-noise, which obeys Poisson statistics and scales linearly with the power incident on the spectrometer. $\sigma_{ex}^2$ is the excess photon noise, in this case due to the pulse-to-pulse fluctuations, or relative intensity noise (RIN), of the SC source. $\sigma_{ex}^2$ scales with the square of the incident power. 
 
 The different dependencies of the variance on the power allows for separation of the three noise sources. The variance should first of all go towards the same spectrometer-determined value $\sigma_{r+d}^2$ for all sources for very low mean counts, which is also confirmed by the experiments. By doing a least-square fit of the measured variance to a linear combination of the 3 individual noise terms, we have quantified their relative strength and plotted them as dashed lines in Figs.~\ref{Fig:NsMsLnSk}(A-C). From Figs.~\ref{Fig:NsMsLnSk}(A-B) we see that for the 80 MHz and 320 MHz SuperK extreme sources, the variance of the noise is dominated by excess noise from the SC source for readings that exceed 250 counts, corresponding to just 6 \% of the 12 bit dynamic range. 
 
 Our results clearly show that the high RIN of the 320 MHz SuperK extreme does not allow the OCT system to be operated in the shot noise limited region, as otherwise reported in \cite{Yun16oOpt}. Here it is important to realize that when the total variance goes from being limited by the read out and dark noise of the detector at low counts [$\sigma_{r+d}^2$ with slope 0 in Figs.~\ref{Fig:NsMsLnSk}(A-C)] to being limited by the excess noise or RIN of the source at high counts [$\sigma_{ex}^2$ with slope 2 in Figs.~\ref{Fig:NsMsLnSk}(A-C)], it will always transition through a certain local region of intermediate count values with slope 1. This can wrongly be interpreted as being shot noise limited if one only locally fits a straight line with slope 1 to the experiments and not surprisingly finds that it fits for a narrow range of count values. One would for example wrongly predict even the 80 MHz SuperK extreme to provide shot noise limited operation by doing so, which it clearly does not. The correct approach is to fit the whole variance versus counts curve to a linear combination of the 3 noise terms as we do here to clearly see their relative strength and be able to determine whether the total variance is shot noise limited or not.
 
In sharp contrast to what was found for the SuperK extreme sources, the excess noise of the ANDi SC, shown in Fig.~\ref{Fig:NsMsLnSk}(C), never dominates within the dynamic range of the spectrometer, and for readings over 1000 counts the recorded variance remains shot-noise limited. Figs.~\ref{Fig:NsMsLnSk} (D-E) show two more representative examples, at 1326 and 1394 nm, comparing the variance of each source. Here, it can be seen that using the ANDi SC source improves the noise characteristics of the OCT system by 12 dB when compared with the 80 MHz SuperK extreme and by 10 dB when compared with the 320 MHz SuperK extreme, for a fixed number of counts of 1100. 

 For a fair comparison of the OCT performance of the three sources we always operate the OCT system at a number of counts that gives the optimum sensitivity performance. For the two SuperK extreme sources this is when the read out and dark noise level equals the excess noise level~\cite{Sor92Sim,Yun03HiSp}, which is at 250 mean counts at 1432 nm [see Fig.~\ref{Fig:NsMsLnSk} (A-B)]. For the ANDi SC source this point is below the shot noise limit and thus cannot be reached. The optimum point of operation is then at the highest number of counts that does not saturate the detector. We therefore chose to operate the ANDi SC source at a fixed number of counts at 1432 nm of 1000.

 To compare the influence of the SC noise on the quality of the A-scans, Fig.~\ref{Fig:NsMsLnSk}(G) shows the variance of the A-scans from the reference arm alone for the three sources, all processed with dark noise subtraction, normalization to the mean of the reference, and spectral apodization using the window shown in Fig.~\ref{Fig:NsMsLnSk}(F). It can be seen that the ANDi SC source provides a significant improvement in A-scan variance for all OPDs of about 13 dB compared to the 80 MHz SuperK extreme and 9 dB compared with the 320 MHz SuperK extreme. This clearly demonstrates that the lower noise level of the ANDi SC source, will result in a much better (darker) background in SD-OCT imaging as compared to what can be obtained with the SuperK extreme sources. \\
The performance of an OCT system is most commonly characterized by its sensitivity, which is directly related to the noise. As we show in the materials and methods section and related Figs.~\ref{Fig:SpecARin}(F-H), the low noise of the ANDi SC source provides a remarkable 12 dB increase in sensitivity compared to the 80 MHz SuperK extreme, which is the direct comparison in terms of repetition rate, and of 7 dB with respect to the 320 MHz SuperK extreme. Our noise and sensitivity characterization of the three SC sources therefore shows that the ANDi SC source provides shot-noise limited SD-OCT imaging with the lowest background noise and highest sensitivity. This is the first clear demonstration of shot-noise limited SC-based SD-OCT operation and, as we shall see in the next section, it translates into a significant improvement in image quality.
\subsection*{Images}
The OCT set up was first characterized for axial resolution using the three sources. The wavelength region from 1280-1478 nm was used in the OCT measurements and was found to have an axial resolution of 5.9 $\mu$m. The lateral resolution was 6 $\mu$m \cite{Niel18Val}. Each of the samples were imaged with the same settings in the OCT system. Identical gray-scaling was ensured for image comparisons. All images have been shadow corrected~\cite{Gir11Shaw} and are displayed on a logarithmic scale. 
\begin{figure}[htbp!]
\centering
\fbox{\includegraphics[width=0.75\linewidth]{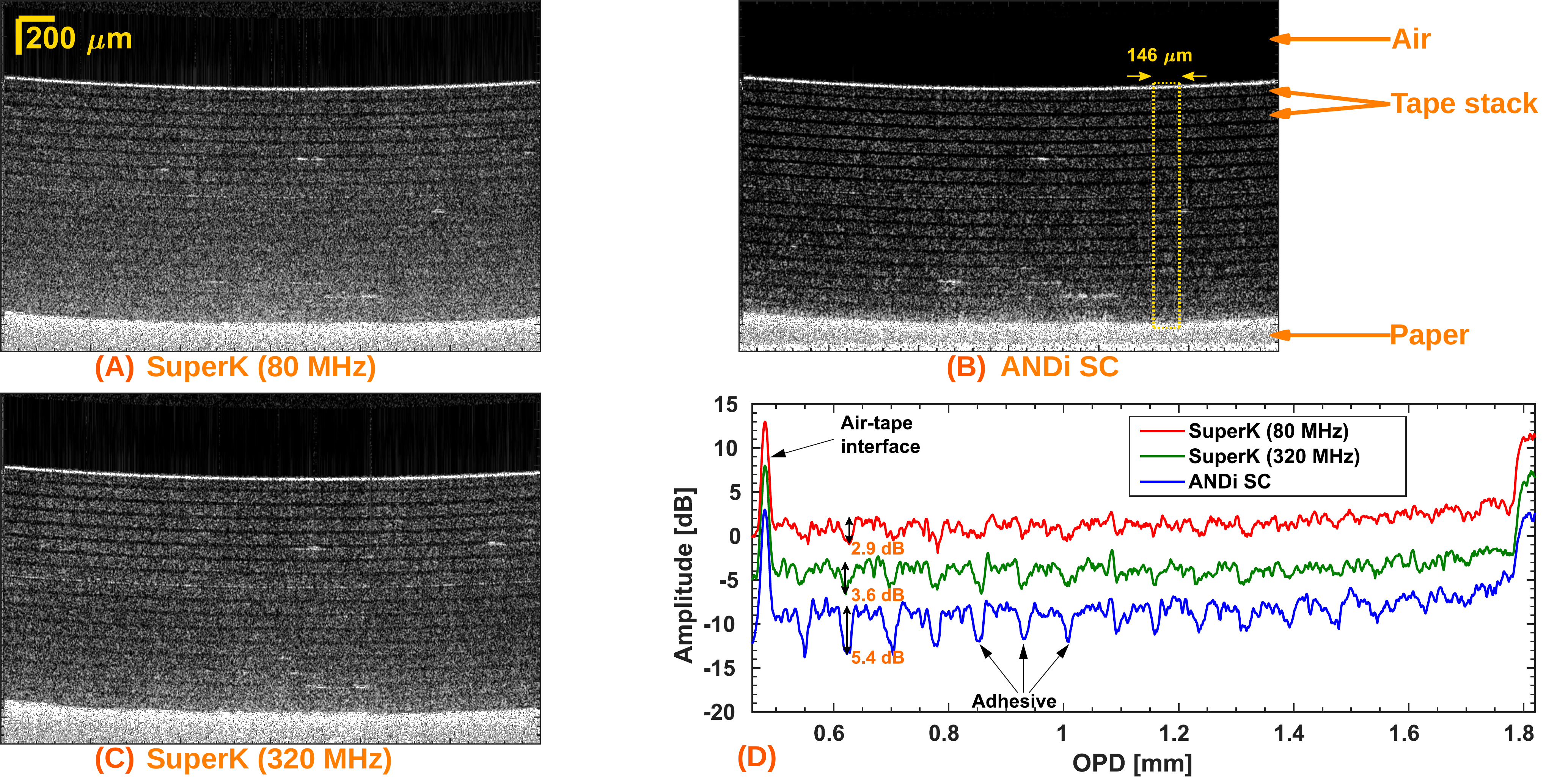}}
\fbox{\includegraphics[width=0.75\linewidth]{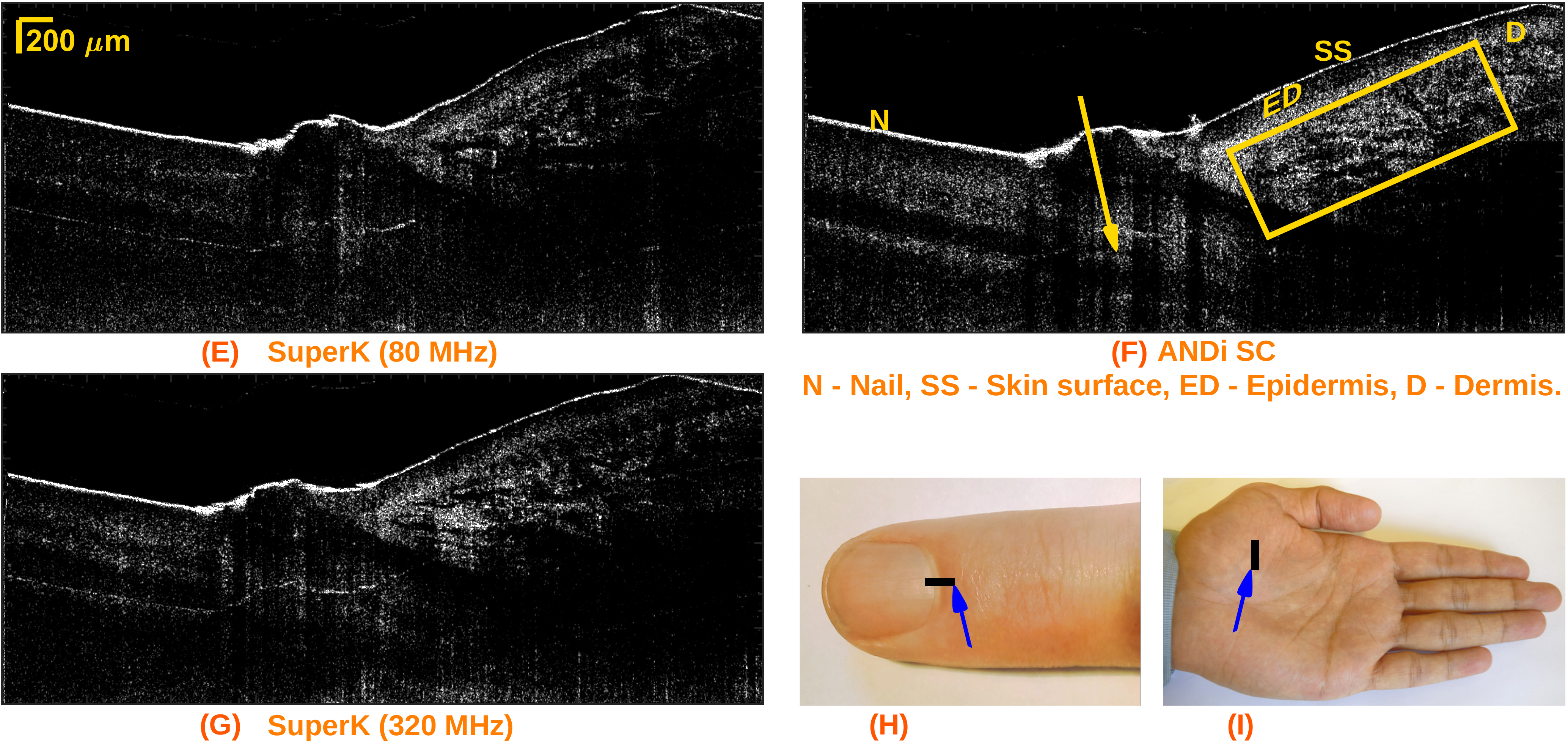}}
\fbox{\includegraphics[width=0.75\linewidth]{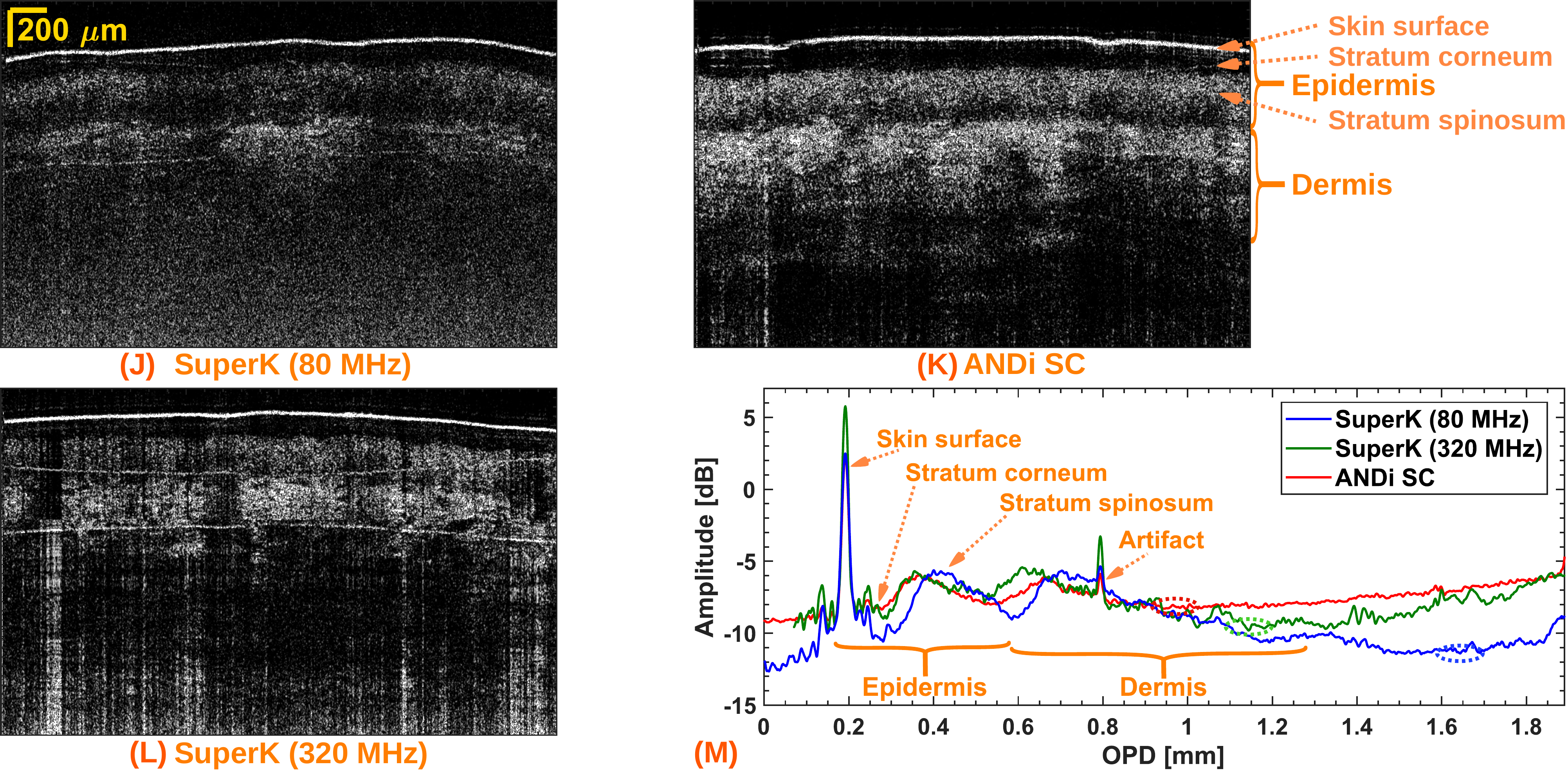}}
\caption{2 $\times$ 3 mm single B-scans of taper layers, obtained using 80 MHz SuperK extreme {\bfseries(A)}, ANDi SC {\bfseries(B)}, and 320 MHz SuperK extreme {\bfseries(C)}. Averaged A-scans within the 146 $\mu$m wide dotted box in \ref{Fig:Imgs}~(B) for the three sources offset by 5 dB {\bfseries(D)}. The average is taken over 49 A-scans equally spaced within the dotted box. 2 $\times$ 4.5 mm single B-scans from a healthy volunteer's finger where the nail begins, obtained using 80 MHz SuperK extreme {\bfseries(E)}, ANDi SC {\bfseries(F)}, and 320 MHz SuperK extreme {\bfseries(G}. Nail fold {\bfseries(H)}. Palm {\bfseries(I)}. 2 $\times$ 3 mm single B-scans from a healthy volunteer's rough skin of palm obtained using 80 MHz SuperK extreme {\bfseries(J)}, ANDi SC {\bfseries(K)}, and 320 MHz SuperK extreme {\bfseries(L)}. Averaged A-scan (averaged over entire B-scan, corresponding to 1024 A-scans) of the rough skin of palm for the three sources {\bfseries(M)}.}
\label{Fig:Imgs}
\end{figure}

In order to easily compare the image quality from each source, we first imaged a stack of 17 layers of tape, each layer of tape consisting of a highly scattering layer of plastic and a thin less scattering layer of adhesive. The single B-scans with an area  of 2 (axial) $\times$ 3 (lateral) mm is shown in Figs.~\ref{Fig:Imgs}(A-C) for the 80 MHz SuperK extreme, ANDi SC, and the 320 MHz SuperK extreme, respectively. The ANDi SC gives the largest signal contrast of the reflections from the plastic to adhesive interfaces as expected. Both the ANDi SC and the 320 MHz SuperK extreme sources enables the OCT system to see through all layers and see the paper substrate, while the 80 MHz SuperK extreme source does not allow to clearly distinguish the bottom interfaces. 
Figure~\ref{Fig:Imgs}(D) shows averaged A-scans within the 146 $\mu$m wide dotted box in Fig.~\ref{Fig:Imgs}(B). The average is taken over 49 A-scans equally spaced with the dotted box. The plastic is scattering the signal much more than the adhesive, and thus the broad maxima in the A-scans represent the tape, while the narrow dips represent the adhesive. %
The amplitude of the A-scans, for the individual sources are offset by 5 dB for clarity. We see from the averaged A-scan using the ANDi SC that the (adhesive) dips are clear until end of the sample and gives a 5.4 dB contrast in the second adhesive layer, as illustrated in Fig.~\ref{Fig:Imgs}(D). In contrast the SuperK extreme sources only give 2.9 and 3.6 dB for the 80 MHz and 320 MHz, respectively. 
It is important to note that in order to assess the pulse-to-pulse noise influence in images, the ANDi SC source should be compared with the 80 MHz SuperK extreme, as they have similar repetition rates and therefore the same degree of averaging. By demonstrating that the  ANDi SC provides a better signal contrast of the images than what is obtained using the 320 MHz SuperK extreme, we show that even the current state-of-the-art in commercial SC sources specifically designed for OCT, provides significantly poorer image qualities. 
\begin{figure}[tb!]
\centering
\fbox{\includegraphics[width=0.90\linewidth]{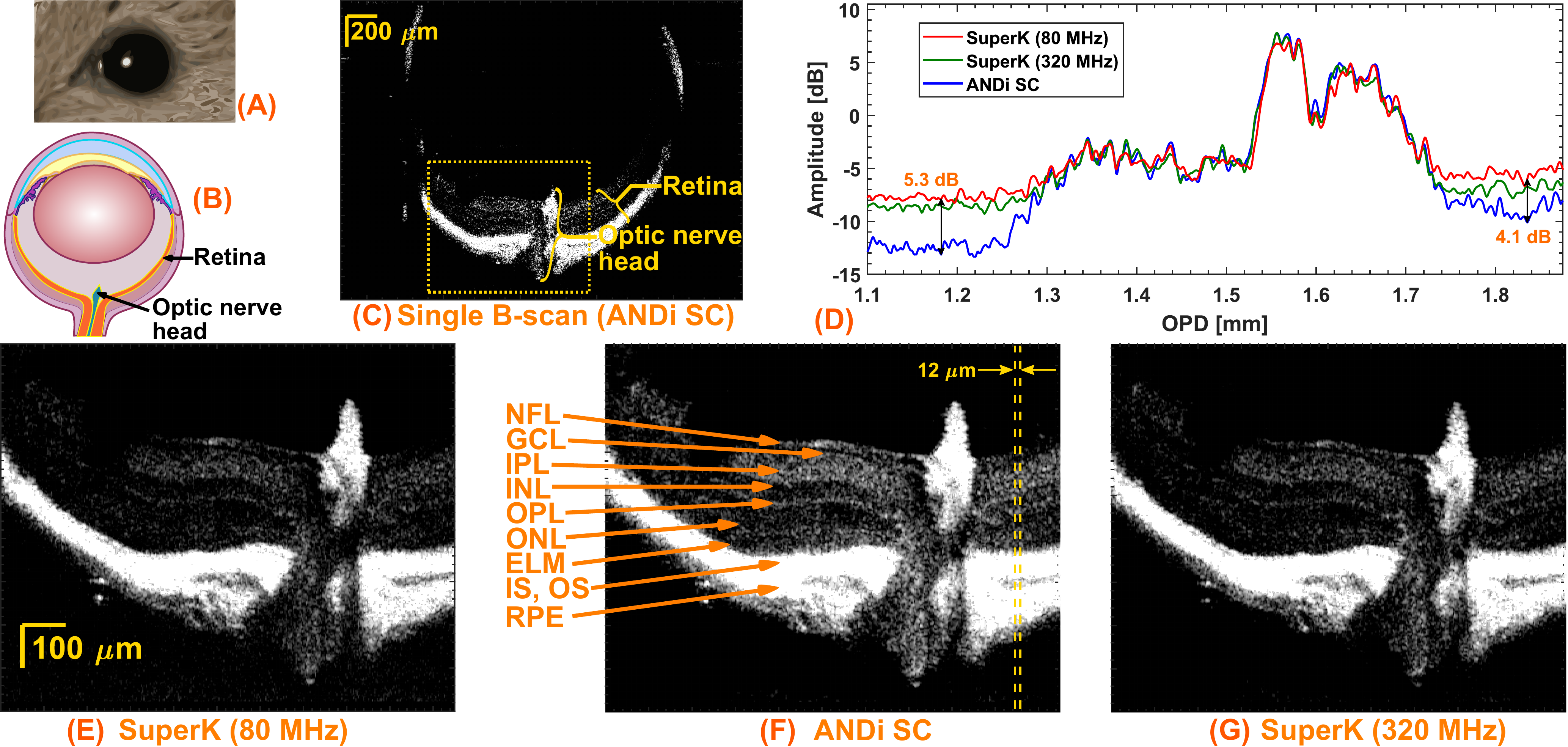}}
\fbox{\includegraphics[width=0.90\linewidth]{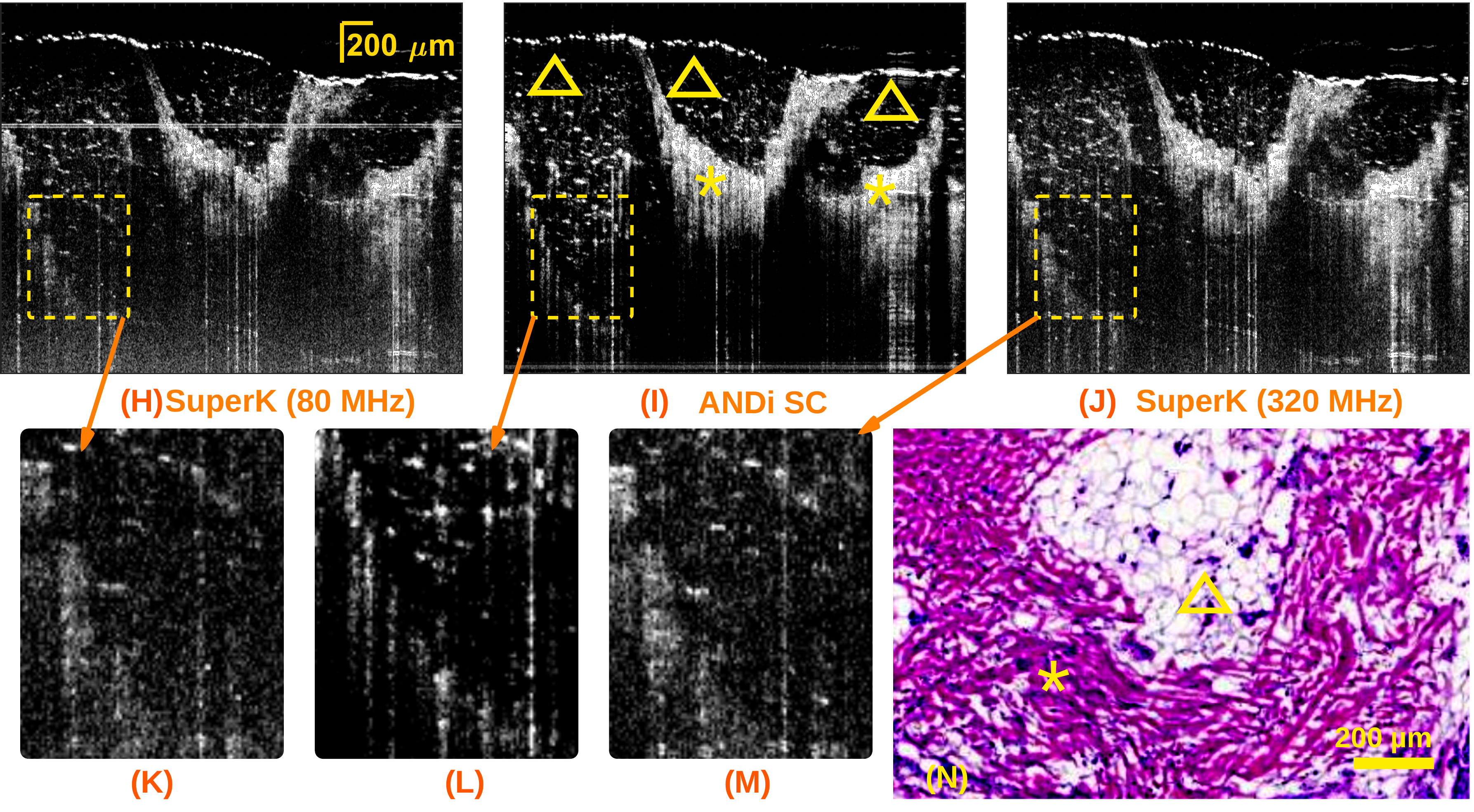}}
\caption{Ex-vivo rat-eye imaging: Artistic image of a mouse eye {\bfseries(A)} with a cross-sectional schematic shown in {\bfseries(B)}. Single B-scan OCT image in depth of a 1.95 $\times$ 2.62 mm section of a mouse retina obtained using the ANDi SC {\bfseries(C)}.Averaged A-scans (averaged over 4 equally spaced A-scans) from the marked region in (F) for the three sources {\bfseries(D)}. Zooms of the retina and optic nerve in the area marked by a yellow box using the 80 MHz SuperK extreme {\bfseries(E)}, ANDi SC {\bfseries(F)}, and 320 MHz SuperK extreme sources {\bfseries(G)} (all averaged over 9 B-scans).
Abbreviations: NFL-nerve fiber layer, GCL-ganglion cell layer, IPL-inner plexiform layer, OPL-outer plexiform layer, ONL-outer nuclear layer, ELM-external limiting membrane, IS, OS-inner segments, outer segments of the photoreceptors, RPE-retinal pigment epithelium. Ex-vivo imaging of human fat tissue: 1.9 $\times$ 3 mm human skin biopsy averaged over nine single B-scans, obtained using 80 MHz SuperK extreme {\bfseries(H)}, ANDi SC {\bfseries(I)}, and the 320 MHz SuperK extreme {\bfseries(J)}. Histology of fat tissue {\bfseries(N)}. Triangles and stars mark tissue types of cutaneous fat tissue and connective tissue of reticular dermis, respectively. }
\label{Fig:ImgsII}
\end{figure}

To study the difference in the images when the OCT system is used to image laterally inhomogeneous biological samples, we first imaged the skin of a healthy volunteer. Images of a 2 $\times$ 4.5 mm section of an index finger nail-fold and a 2 $\times$ 3 mm section of a palm are shown in Figs.~\ref{Fig:Imgs}(E-G) and~(J-L), respectively. Using a marker is was assured that the nail-fold images were taken from the same point, so that a quantitative comparison is justified. Comparing the nail-fold images obtained when using the ANDi SC and the 80 MHz SuperK extreme, it can be seen that the region around and below the nail, as it goes into the finger (arrows), is clearly visible in the image with the ANDi source, but not visible in the image obtained by using the 80 MHz SuperK extreme and only barely visible with the 320 MHz SuperK extreme. In addition, skin in the right side of the image (rectangle), has higher contrast for the ANDi SC and larger imaging depth. .  

To quantify the increased penetration, we imaged the palm, which has a more regular surface than the nail fold, see Fig.~\ref{Fig:Imgs}(J-L). For the palm the imaged area is only approximately the same and while the images with the 80 MHz SuperK extreme and the ANDi source are taken shortly after one another, the image with the 320 MHz SuperK extreme is taken several days later. This means that the images can only be qualitatively compared. As with the nail fold, the ANDi SC shows darker background, higher contrast, and improved penetration. 
Fig.~\ref{Fig:Imgs}(M) shows A-scans for each source, averaged across the entire B-scan (1024 A-scans). The curves are shifted horizontally and vertically such that the major peak from the air-skin interface overlap, allowing a relative comparison of the signals from within the skin. We see that the epidermis and dermis signals (indicated local broad maxima) are of similar strength with all three sources, but that the dark part in the stratum corneum is darker when the ANDi SC is used. Also, the 80 MHz and 320 MHz SuperK extreme curve hit the noise floor at a penetration depth of 0.78 mm and 0.96 mm below the surface, respectively, whereas the ANDi SC reaches the noise at 1.45 mm below the surface. The measured penetration depths (marked on the curves and in the images) demonstrate that the ANDi source provides an increased penetration depth of about 0.5 mm even compared to the 320 MHz SuperK extreme.

To demonstrate the image quality improvement within ophthalmology, we imaged a mouse retina sealed in epon resine~\cite{Bar19MiR}, with each of the three sources. A single B-scan image of the full mouse retina, obtained using the ANDi SC, is shown in Fig.~\ref{Fig:ImgsII}(C). In Figs.~\ref{Fig:ImgsII}(E-G) we zoom on the lower part of the retina around the optic nerve, corresponding to the region in Fig.~\ref{Fig:ImgsII}(C) marked with a dashed box, and show averaged B-scans (averaged over 9 B-scans) to better highlight individual features, such as layers of different cell types and the optic nerve head. The various cell layers are labelled in Fig.~\ref{Fig:ImgsII}(F). Since the retina is thin and transparent, all the three sources were able to penetrate the entire retina, but the ANDi SC source provides a much better contrast. The improvement in the contrast is up to 5.3 dB, as can be seen quantitatively from the averaged A-scans for the three sources shown in Fig.~\ref{Fig:ImgsII}(D), obtained by averaging across the 12 $\mu$m region within the two dashed lines in Fig.~\ref{Fig:ImgsII}(F). 

To further demonstrate the sensitivity enhancement obtained by using the ANDi SC source, we image a another type of tissue being that of cutaneous fat tissue found in the dermis of the skin. A biopsy was extracted according to the procedure of Mohs surgery. After defrosting the biopsy it was imaged with the three different sources. The result (averaged over nine individual B-scans) are seen in Fig.~\ref{Fig:ImgsII}(H-J). Due to the already demonstrated better contrast the ANDi SC source provides a significantly improved information content, which is especially observed in the deeper parts of the tissue. From the zooms presented in Figs.~\ref{Fig:ImgsII}(K-M) we see for example that the scattering points from the hexagonal-like structures of the cutaneous fat tissue [marked by triangles in Fig.~\ref{Fig:ImgsII}(H-J) and the histology image in Fig.~\ref{Fig:ImgsII}(N)], are much more clearly recognized with the ANDi SC source. 

In general, the quality of the images obtained with the 320 MHz SuperK extreme is better than the ones obtained using the 80 MHz SuperK extreme, due to the higher repetition rate and thereby increased averaging. However, even with 3.6 times lower repetition rate the ANDi SC performs better than the 320 MHz SuperK extreme because of its extraordinary low-noise properties.

\section*{DISCUSSION}
\subsection*{Achievable sensitivity}
\begin{figure}[htbp!]
\centering
\fbox{\includegraphics[width=0.6\linewidth]{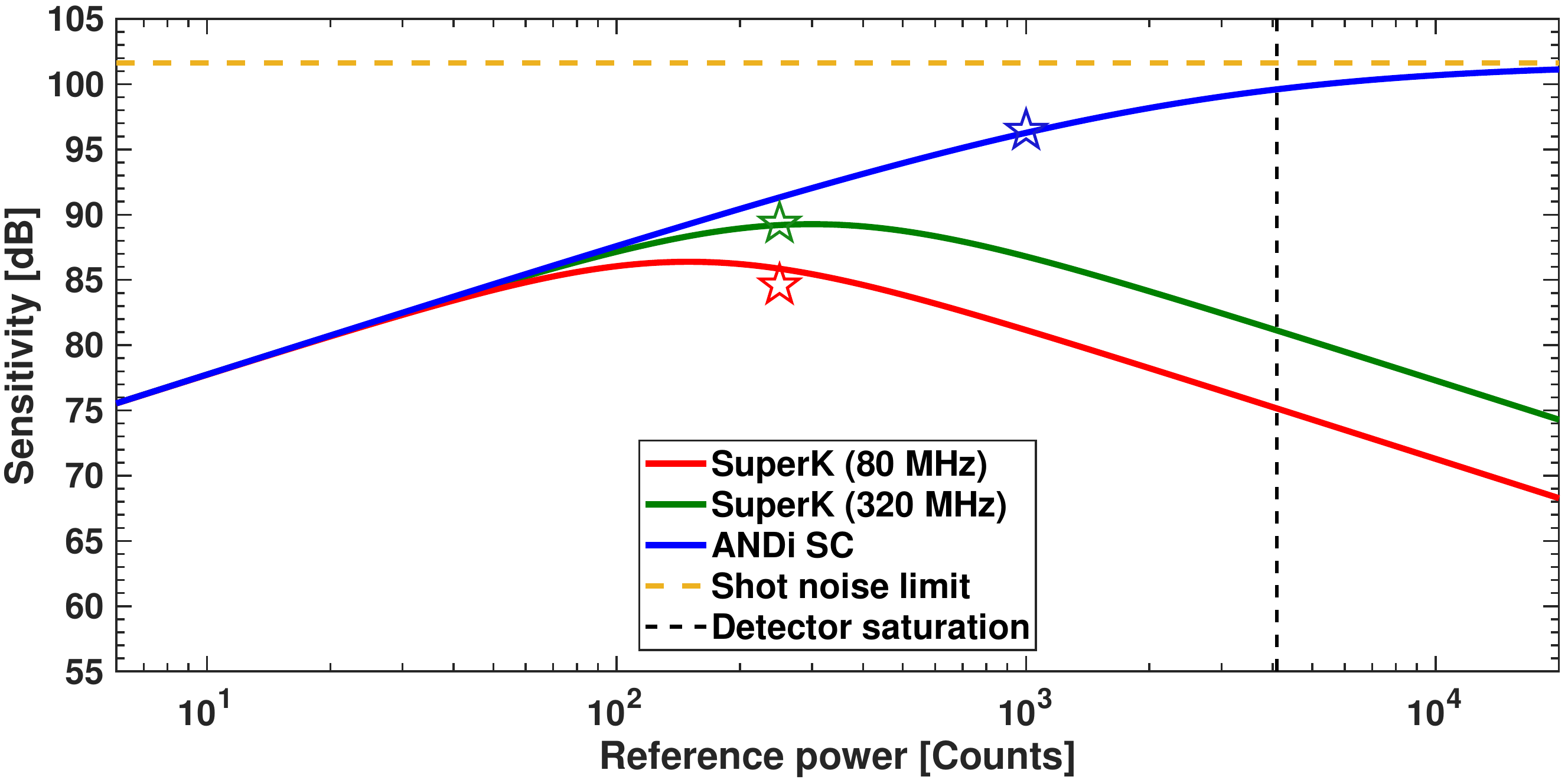}}
\caption{Theoretical sensitivity as a function of reference power for the 80 MHz SuperK extreme, 320 MHz SuperK extreme, and ANDi SC sources. Stars represents the experimentally obtained sensitivities presented in the Materials and Methods section. Model parameters are found from experimental system characterization and are given in detail also in the Materials and Methods section. Calculations are based on \cite{Leit03FdTd,Jen19Noi}.}
\label{Fig:SensTheory}
\end{figure}

In order to assess the sensitivity values produced by the three source configurations for arbitrary values of the reference power, we have modelled the sensitivity theoretically under conditions of the experiments. The theoretical description of the OCT sensitivity is well-known and is commonly in short form expressed as the signal-to-noise ratio (SNR) for a perfect reflector~\cite{Leit03FdTd}: Sensitivity $= SNR_{max} = \tilde{S}_{max}^2/\big(\tilde{\sigma}^2_{r+d}+\tilde{\sigma}^2_{shot}+\tilde{\sigma}^2_{ex}\big)$ with all components now evaluated in the OPD domain, signified by a tilde. Here $\tilde{S}_{max}$ is the maximum of the detected point spread function of the reflector and the sigmas are the respective noise contributions from spectrometer, shot noise and excess noise stemming from the SC source. The traditional representation of the RIN (and therefore $\tilde{\sigma}^2_{ex}$) as being that for a thermal source has been shown to overshoot the RIN of a supercontinuum source and we thus apply a more precise measurement-based model~\cite{Jen19Noi}. With this, we present the theoretically calculated sensitivities for the three sources in Fig.~\ref{Fig:SensTheory} as a function of power returned from the reference arm. The sensitivities for the experimentally used optimum (for our systems) reference power settings are marked by stars.

From Fig.~\ref{Fig:SensTheory}, we find the expected behaviour of the sensitivity for the three different sources. The 80 MHz and 320 MHz SuperK extreme sources must be operated at the reference power, at which the total noise becomes dominated by the RIN of the source, which limits the maximum sensitivity. In contrast, when applying the ANDi SC source, the reference power can be increased to a level close to detector saturation in order to approach the shot-noise limit of maximum sensitivity, which is 102 dB for our system (yellow dashed line in  Fig.~\ref{Fig:SensTheory}). Due to the high read out and dark noise of the InGaAs (indium gallium arsenide) detectors this reference power level is at around 1000 counts, at which the sensitivity is 96 dB. This is 7 dB above the maximum achievable sensitivity when using the 320 MHz SuperK extreme source, but still 6 dB below the maximum shot-noise limit. 

The sensitivity level of an OCT system depends on several factors, such as power, integration time, the quality of the detector, and the optical loss from the sample to the detector. A better detector would allow to operate closer to the shot-noise limit and thereby improve the sensitivity. Furthermore, the loss of a broadband fiber-based SD-OCT system as ours is rather high, mainly due to losses in the coupling to and from the fibers and the fiber coupler. A characterization showed that only a fraction of 6\% of the power returned from the sample reached the spectrometer. A reduction of the loss would significant improve the obtainable shot-noise limited sensitivity. A realistic improvement of the fraction of returned power of for example a factor of two would intrinsically lift the achievable shot-noise limit to 105 dB. 

With this in mind it is interesting to discuss the work from 2016 by Yuan et al., in which they reported the current record sensitivity of 107 dB for SC-based SD-OCT, when using the commercial 320 MHz SuperK extreme source from NKT Photonics, with similar sample power and spectrometer integration time and depth resolution as we use here \cite{Yun16oOpt}. The sensitivity result of theirs is thus directly comparable to those obtained with the 320 MHz SuperK extreme source presented in this paper. The main difference is that in \cite{Yun16oOpt} the OCT system was operated at 800 nm with a line-scan CCD camera, so both the detector and the optical system is different from our system. However, assuming that the relative improvement of 7 dB when going from 320 MHz SuperK extreme to ANDi SC still holds, this  would translate our ANDi SC result to a record 114 dB sensitivity using their 800 nm SC-based OCT system. 

The record sensitivity of 96 dB achieved here with the ANDi SC-based 1300 nm OCT system and potential 114 dB achievable at 800 nm, is competing with the leading OCT technology in terms of sensitivity, which is swept source OCT (SS-OCT). SS-OCT is a technology that in most cases musters a 105-110 dB OCT sensitivity but is however limited to a spectral bandwidth of about 150 nm in, e.g., the 1300 nm range \cite{Hub05Cmd,Hub06Fdml,Wie10Mul}, which means that it has a limited axial resolution of about 7 µm. Our results predict a future ANDi SC-based SD-OCT system, which has the ultra-high axial resolution of SD-OCT systems and the ultra-high sensitivity of SS-OCT systems.

\subsection*{Conclusion}
Supercontinuum research has unlocked an unprecedented axial resolution in OCT. The gist of the technology being that the axial resolution is fundamentally decoupled from the focus choice of the scanning beam and instead determined by the spectral bandwidth of the source. 

In this work we have shown that an ANDi SC source, in which the SC is generated using an ANDi fiber pumped by a short fs pulse, can have so low-noise that it enables excess noise-free SD-OCT. We have compared the noise properties of a high resolution SD-OCT system operating around 1370 nm using an ANDi based low-noise SC source to the noise properties of the system when two commercially available SC sources displaying high noise are used. Using the ANDi SC source, we have for the first time demonstrated that SD-OCT can now be operated in the shot-noise limited detection regime. With the improved noise characteristics, we find an OCT sensitivity gain of 12 dB is inherently won, while maintaining an axial resolution of 5.9 µm and with only a slight increase of repetition rate from 80 MHz to 90 MHz, as presented in Figs.~\ref{Fig:NsMsLnSk}(D and G). 

The value of this improvement is first of all manifested in the increased contrast of the retinal layers of the eye against the background, in this work represented by an ex-vivo mouse retinal sample depicted in Fig.~\ref{Fig:ImgsII}(A-C). The lower background noise enables faster imaging without significant loss of image quality. With OCT known for revolutionizing the ophthalmological diagnosis, the performance of delineating the retinal cell layers to a high precision is decisive and this demands both best possible performance in axial resolution and sensitivity, when tracking individual cell layers. 

In dermatology, OCT has proved to increase the diagnostic performance of skin cancer achieving an accuracy for basal cell carcinoma of $\sim 60\%$ and in melanoma of $\sim 40\%$~\cite{Ruf18Oct}. Finding the transition between malignant and benign tissue and dismissing cancer tissue emitators in OCT images is however still very challenging. By imaging nail fold, hand palm and ex-vivo subcutaneous fat tissue, we documented a substantial improvement in cell structure transitions as well as increased penetration depth, the latter being ever important in the pursue of recognizing and removing the full extension of malignant tissue. 

In general, noise properties and the images obtained by using 80 MHz and 320 MHz SuperK extreme show that having a high repetition rate improves the details in the image, when the sources have high pulse-to-pulse fluctuation. The ANDi fiber based low-noise source, on the other hand, does not rely on pulse averaging to keep the noise low and is therefore a prime candidate to take advantage of the ever-increasing spectrometer speeds without sacrificing image quality. 
\section*{MATERIALS AND METHODS}\label{MandM}
\subsection*{Experimental setup}
\begin{figure}[tb!]
\centering
\fbox{\includegraphics[width=0.47\linewidth]{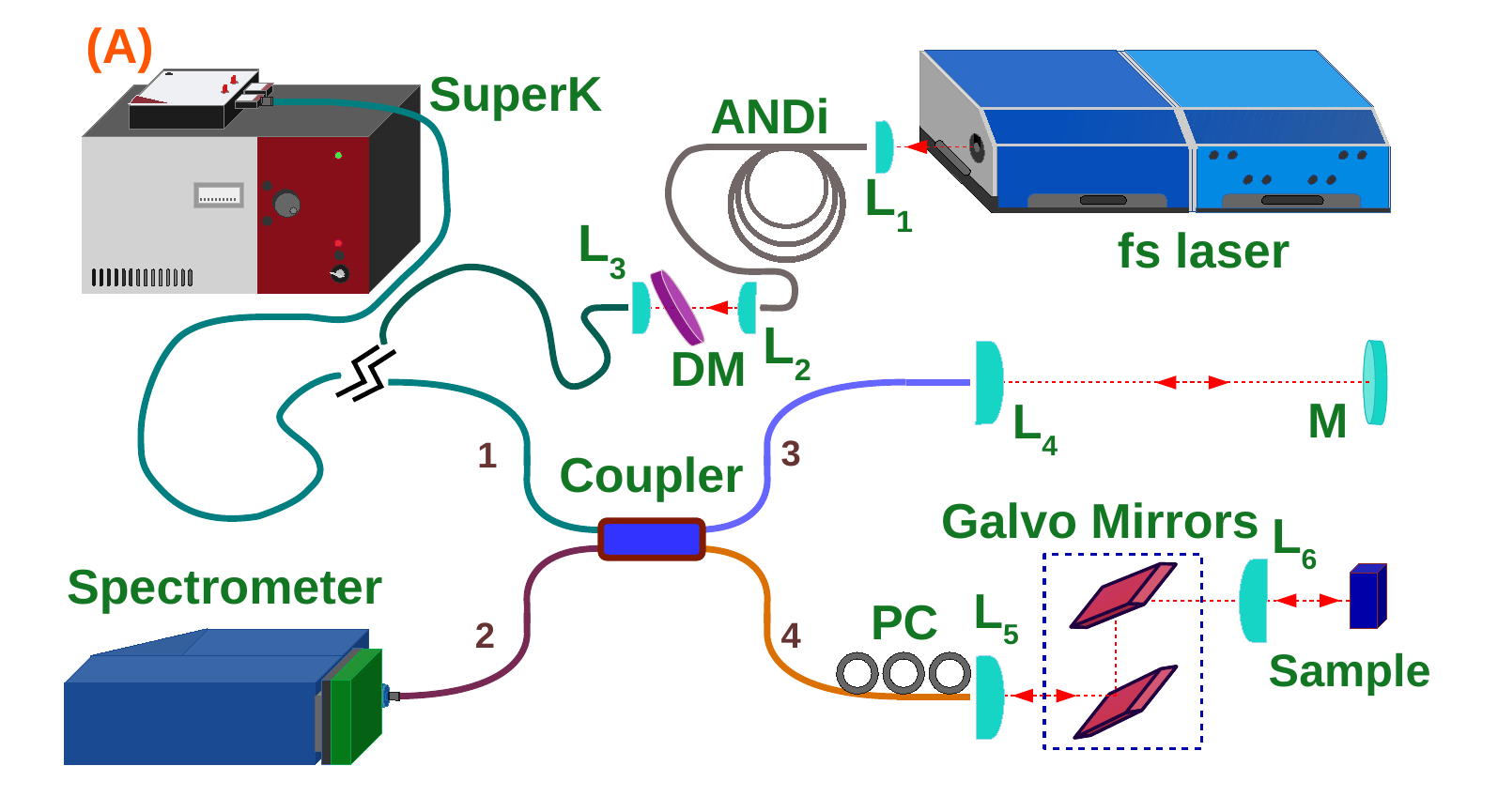}}
\fbox{\includegraphics[width=0.47\linewidth]{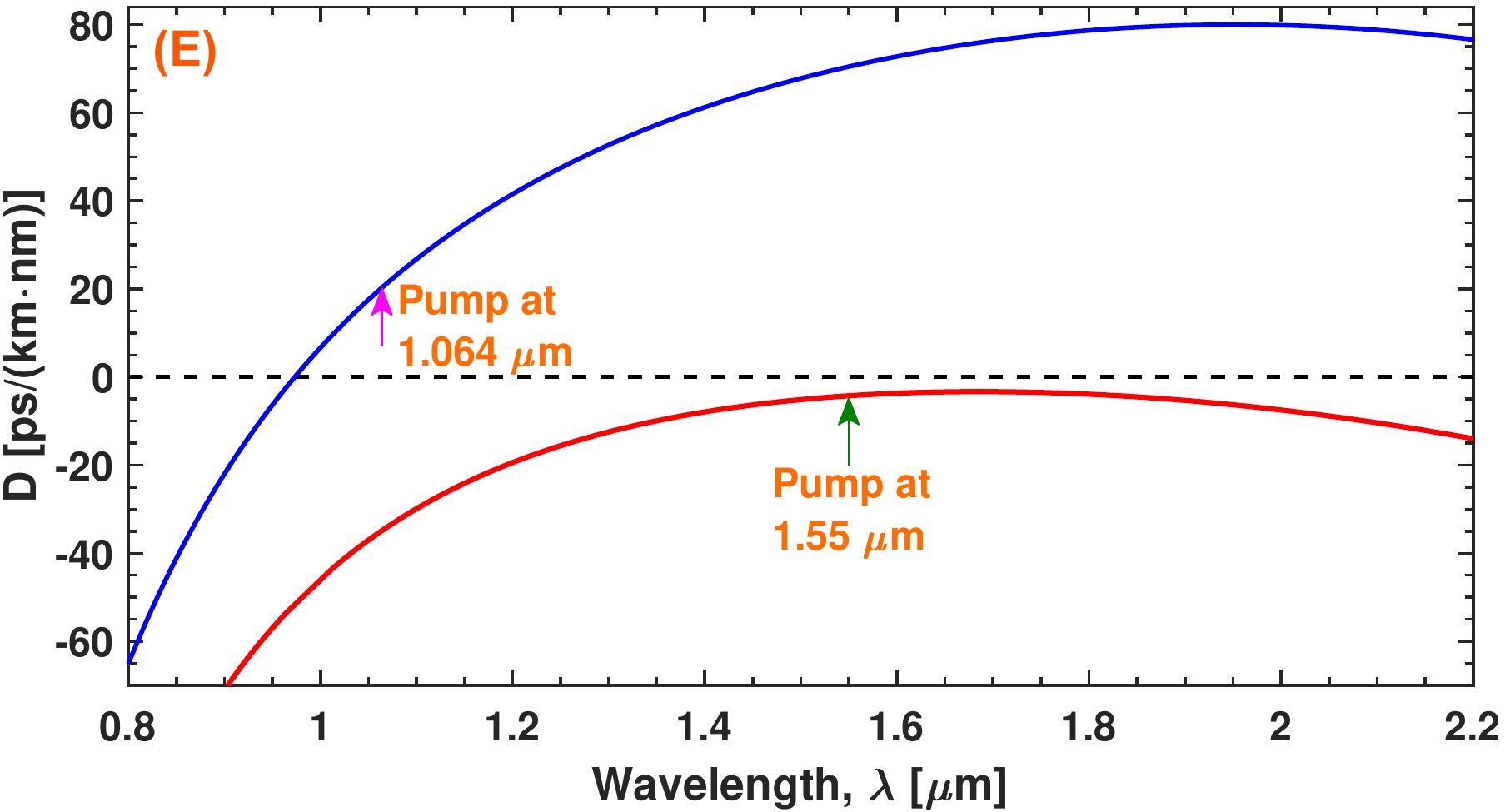}}
\fbox{\includegraphics[width=0.47\linewidth]{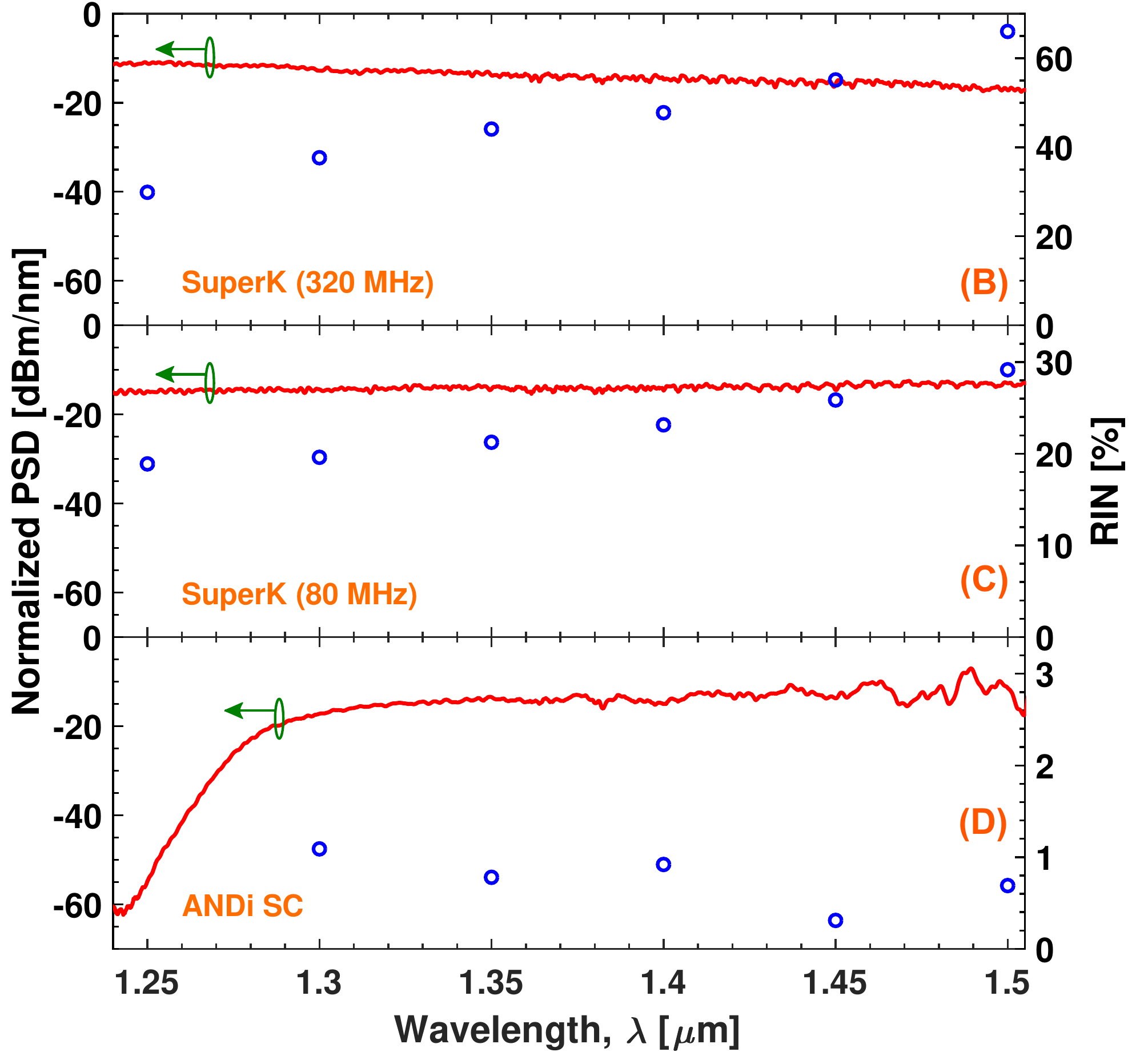}}
\fbox{\includegraphics[width=0.47\linewidth]{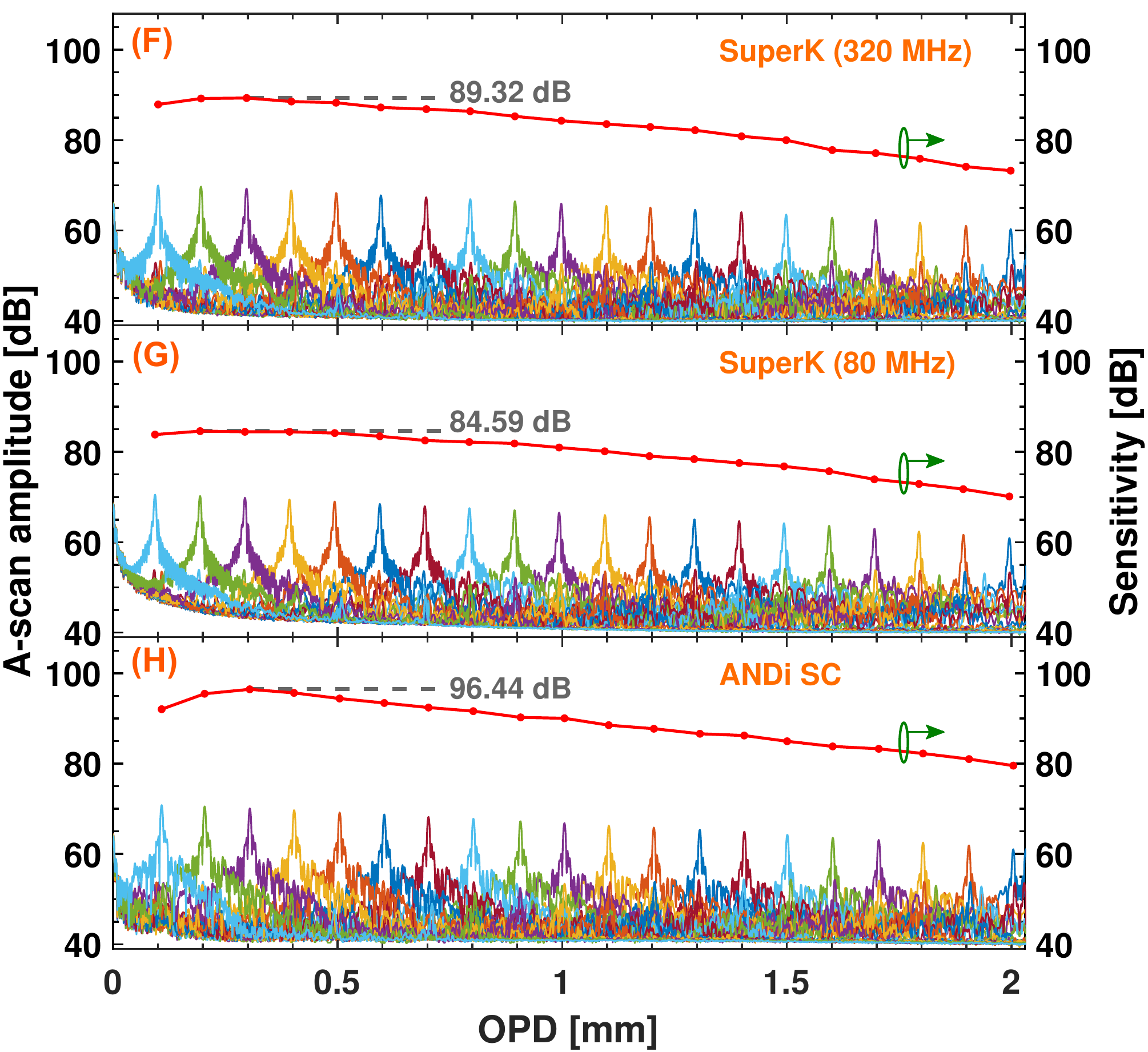}}
\caption{Schematic of the OCT set up {\bfseries(A)} [L$_i$: lens, M: mirror, DM: dichroic mirror, and PC: polarization controller]. Normalized power spectral density (PSD) profiles measured using an OSA (red curves, left axis) and RIN profiles (blue circles, right axis) for the 320 MHz SuperK extreme {\bfseries(B)}, the 80 MHz SuperK extreme {\bfseries(C)}, and the ANDi based low-noise SC source {\bfseries(D)}. Calculated dispersion profile for the nonlinear fiber used in the SuperK extreme (blue) and the measured dispersion of the GeO$_2$ doped silica ANDi fiber (red) {\bfseries(E)}. Mean A-scan (over 1024 measurements) of a sample mirror at 20 different axial positions on the left axis, and the sensitivity on the right, for the SuperK extreme at 320 MHz {\bfseries(F)} and 80 MHz {\bfseries(G)}, and the ANDi based low-noise SC source {\bfseries(H)}.}
\label{Fig:SpecARin}
\end{figure}

The SD-OCT setup, shown in Fig.~\ref{Fig:SpecARin}(A), is a Michelson interferometer with an ultra-broadband 50/50 coupler (Thorlabs: TW1300R5A2) splitting the light into a reference arm (arm labelled `3') and a sample arm (arm labelled `4'). The sample arm contains a two-axis galvo scanner (Thorlabs: GVS002) after which the light is focussed onto the sample using a lens. The spectrometer in the arm labelled `2' has a detection range from 1074-1478 nm with a resolution of 0.2 nm (Wasatch Photonics: C-1070-1470-GL2KL). The spectrometer has a 12 bit quantization for the amplitude and 2048 pixels. In the source arm `1' three different sources were used. For the low-noise ANDi source, a 125 fs full width at half maximum  pump at 1.55 $\mu$m (Toptica: Femto fiber) operating at 90 MHz repetition rate was coupled into 10 m of GeO$_2$ doped silica fiber (OFS Denamrk) using an aspheric lens. The fiber has an ANDi profile. The dispersion of the fiber was measured from 0.86 to 2.38 $\mu$m using white light interferometry and is plotted in  Fig.~\ref{Fig:SpecARin}(E) (red). The SC generated in the ANDi fiber spans from 1280-1910 nm at the $-$30 dB level.

The light out of the ANDi fiber was collimated using an aspheric lens and a dichroic mirror (Thorlabs: DMSP1500) labelled `DM' was used to cut off light above 1500 nm. The light was then free space coupled into arm `1' of the interferometer using an aspheric lens. The other two sources used were a SuperK extreme operating at a repetition rate of 80 MHz (NKT Photonics: SuperK EXTREME EXW-12) and a SuperK extreme operating at a repetition rate of 320 MHz (NKT Photonics: SuperK EXTREME EXR-9 OCT) developed specifically for OCT. The calculated dispersion of the nonlinear fiber in these sources is plotted in Fig.~\ref{Fig:SpecARin}(E) (blue). The SC was spectrally filtered using a long pass filter at 1.0 $\mu$m (Thorlabs: FELH1000) and a short pass filter at 1.5 $\mu$m (Thorlabs: DMSP1500) and fiber coupled to arm `1' of the interferometer. For direct comparison of the images obtained using the three sources, in each of the OCT measurements it was ensured that the average power in the sample arm was $\sim$3.5 mW. Integration time of the spectrometer for all of the measurements was 9.1 µs and the spectral line scan rate was 76 kHz.
\subsection*{Spectra and RIN measurement} 
The spectra used for the OCT measurements are shown in Fig.~\ref{Fig:SpecARin} (B-D, red, left axis). When a fs pulse is used to pump a fiber with a weak ANDi profile the spectral broadening is dominated by SPM and OWB and these broadening processes can lead to a coherent spectrum. For the SC from the SuperK extremes the nonlinear fiber has a ZDW around 0.98 $\mu$m, as shown in  Fig.~\ref{Fig:SpecARin} (E, blue), and a ps pump centered at 1.064 $\mu$m is used. The spectral broadening is dominated by modulation instability, which generates a sea of solitons in the anomalous dispersion regime, which then interact and collide while red-shifting due to the Raman-induced soliton self-frequency shift. In this case the noise property of the generated spectra is dominated by the amplification of quantum noise and subsequent phase- and amplitude-dependent soliton collisions, which make the spectra incoherent. Even though the spectra measured by the optical spectrum analyser (OSA) (Yokogawa: AQ6317B) show a flat profile in Fig.~\ref{Fig:SpecARin}(B and C), there is, in fact considerable spectral fluctuation from pulse-to-pulse. In order to quantity these fluctuations, we performed spectrally resolved pulse-to-pulse RIN measurements~\cite{Laf09Dir,Shr19Uln}. The spectrally resolved RIN measurements were performed by using several 10 nm bandpass filters in intervals of 50 nm. The bandpass filtered SC was coupled into a 50 $\mu$m core diameter fiber. This was then fiber coupled into a 5 GHz bandwidth InGaAs detector with an active area diameter of 80 $\mu$m (Thorlabs: DET08CFC). The pulse train out of the photodiode was recorded with a fast oscilloscope (Teledyne: LeCroy-HDO9404, 4 GHz bandwidth, 40 GS/s). 
The linear region of the detector was initially determined by using the bandpass filtered spectrum from the pump. During the measurement of the RIN of the SC, it was ensured that the detector was being operated in this linear regime by suitably adjusting the power into the detector.  The peaks of the measured trace, which are proportional to the pulse energy, were extracted and used to find the RIN$=\sigma/\mu$, where $\sigma$ is the standard deviation, and $\mu$ is the mean. For the ANDi fiber based SC source 45,135 pulses were used for each measured RIN value and the measured RIN is shown in Fig.~\ref{Fig:SpecARin}(D, right axis). The excellent noise properties of the ANDi fiber based source is confirmed by the low RIN values, which is below 1.1\% in the entire wavelength range. For the 80 MHz SuperK extreme, 249,078 pulses were used for each measured RIN value ad the measured RIN is shown in. Fig.~\ref{Fig:SpecARin}(C, right axis). The RIN is 18.9\% at 1.25 $\mu$m and increases to 29.2\% at 1.5 $\mu$m. 
Similarly, for the 320 MHz SuperK extreme, 997,288 pulses were used for each measured RIN value ad the measured RIN is shown in Fig.~\ref{Fig:SpecARin}(B, right axis). The RIN is 29.9\% at 1.25 $\mu$m and increases to 66\% at 1.5 $\mu$m. The 320 MHz SuperK extreme has higher pulse-to-pule RIN than the one operating at 80 MHz. However, in the OCT set up the spectrometer records lower noise for the 320 MHz SuperK extreme [see Figs.~\ref{Fig:NsMsLnSk}(A-B and D-E)] as the spectrometer averages over time and the 320 MHz SuperK extreme has four times more pulses than the 80 MHz SuperK extreme. These measurements show that the ANDi based SC source has very low pulse-to-pulse RIN, more than an order of magnitude lower than the commercially available SC sources.
\subsection*{Characterization of the OCT setup} 
The axial resolution of the OCT system was determined to be 5.9 µm, averaged over all OPDs. No significant broadening was observed at larger OPD (when the ANDi SC was used for e.g., the axial resolution was 5.77 $\mu$m at 0.5 mm, which increased to 5.96 $\mu$m at 1.8 mm). The lateral resolution was measured to 6 µm \cite{Niel18Val}. In OCT, the sensitivity is defined as the minimum sample reflectivity, which gives a signal-to-noise ratio (SNR) of 1. The sensitivity is calculated as $SNR_{max} = S_{max}^2/\sigma_{tot}^2$~\cite{Agr17Met}. After the reference power was adjusted such that the measured photo-counts alone from the reference arm  was set at the optimum point; that being 250 for the SuperK extremes, a mirror was placed in the sample arm. The reflection of the full power would saturate the spectrometer, so the sample mirror was slightly misaligned to attenuate the power collected by the OCT system below the saturation limit. This attenuation was added to the signal/noise fraction to obtain the sensitivity. The sensitivity roll-off represents the signal degradation with the OPD due to both finite pixel size and spectral resolution in the spectrometer. The sensitivity roll-off of the OCT system when using the 320 MHz SuperK extreme was measured to be 9.5 dB/mm, 8.0 dB/mm when using the 80 MHz SuperK extreme, and 8.8 dB/mm when the ANDi SC source was used.
\subsection*{Image processing} 
For imaging, the sample was positioned in the negative OPD such that the deeper layers were closer to OPD = 0 mm. This was to have the sensitivity roll-off work against the scattering of the sample, such that the deeper layers with weak signal were in the low OPD region with high sensitivity. All images are shadow corrected using the approach of Girard et al.~\cite{Gir11Shaw}, and they are plotted on a logarithmic scale. A Hamming window was used in the range 1280 to 1478 nm, and zeros from 1074-1280 nm. The value provided for the axial depth for each of the image is the corresponding OPD. In addition to shadow correction, only for the image with tape-layer phantom, the data before the main peak was smoothened by averaging.
\subsection*{Calculation of theoretical sensitivity curves} 
The values used for the theoretical sensitivity curves, plotted in Fig.~\ref{Fig:SensTheory}, are based on the equation for sensitivity denoted as $\Sigma_{FDOCT}$ in Ref.~\cite{Leit03FdTd} with relative intensity noise extension of \cite{Jen19Noi}, is provided in the following. Fraction of the total power from sample arm exiting the interferometer is $\gamma_s= $0.063 (measured), fraction of the total power from reference arm exiting the interferometer is $\gamma_r= $0.133 (measured), spectrometer efficiency (losses in spectrometer before hitting the detector)-$\rho=$0.5 (from vendor), quantum efficiency of the detector-$\eta=$0.7 (from vendor), exposure time of the detector-$\tau=$9.2 $\mu$s, center wavelength applied is $\lambda_c=1370$, polarization degree is assumed to be zero, power associated with center frequency (measured sample and reference power in total)-P$_{0}=$10.4 mW, spectrometer gain factor $\Delta e=270$ electrons/count and receiver noise $\sigma_{receiver}=810$ electrons, the latter two specified by the vendor. 

\printbibliography[title={REFERENCES AND NOTES}]
\setlength\parindent{0pt}\vspace{1ex}
{\small \textbf{Acknowledgements:} OFS Denmark is acknowledged for providing the ANDi fiber. Dr. \'{A}lvaro Barroso, Medical Faculty University of M\"{u}nster, Germany, is acknowledged for his technical inputs during the sample selection process of ex-vivo mouse retina. \textbf{Funding:} European Union's Horizon 2020 research and innovation program under the Marie Sk\l{}odowska-Curie grant agreement No. 722380 (the project SUPUVIR); Innovation Fund Denmark under the project J. No. 4107-00011A (the project ShapeOCT). Det Frie Forskningsr\aa{}d (DFF) under the project No. LOISE-4184-00532B (the project LOISE).} \textbf{Author contributions:} S.R.D.S. developed the ANDi SC and characterized the ANDi fiber and RIN of all the three sources. L.G-N .advised on the ANDi fiber selection. J.T.O. provided the 80 MHz SuperK extreme and discussed the project. N.M.I built the SD-OCT setup. S.R.D.S., M. K., and N.M.I. characterized the OCT setup and performed OCT on the samples, and processed the images. P.S., J.K., and J.S. provided the mouse retina sealed in epon resine and advised on its imaging. M.G. and M.M. obtained the fat tissue sample and advised on its imaging. N.M.I prepared the fat tissue for imaging. O.B. supervised the development of the ANDi SC and the OCT imaging and initiated and guided the project. S.R.D.S., M.J., N.M.I, and O.B. drafted the paper. All the authors discussed the manuscript. \textbf{Competing interests:} The authors declare that they have no competing interests. \textbf{Data and materials availability:} All data needed to evaluate the conclusions in the paper are present in the paper. Additional data related to this paper may be requested from the authors.
\end{document}